\documentclass[final]{aipproc}
\pdfoutput=1

\layoutstyle{6x9}

\begin{document}

\title{ Higher-dimensional Higgs Representations in SGUT models
}




\author{Alfredo Aranda}{
  address={Facultad de Ciencias,CUICBAS,  Universidad de Colima,
  Bernal D\'{i}az del Castillo 340, Colima, Colima, M\'exico },
altaddress={Dual C-P Institute of High Energy Physics, M\'exico},
}

\author{J. L. D\'iaz-Cruz}{
  address={C.A. de Part\'iculas, Campos y Relatividad ,
  FCFM-BUAP, Puebla, Pue. M\'exico},
   altaddress={Dual C-P Institute of High Energy Physics, M\'exico},
}

\author{Alma D. Rojas}{
address={Facultad de Ciencias,CUICBAS,  Universidad de Colima,
  Bernal D\'{i}az del Castillo 340, Colima, Colima, M\'exico },
}

\begin{abstract}

Supersymmetric Grand Unified Theories (SGUTs) have achieved some
degree of success, already present in the minimal models (with
SU(5) or SO(10)). However, there are open problems that suggest
the need to incorporate more elaborate constructions, specifically
the use of higher-dimensional representations in the Higgs sector.
For example, a $45$ representation of SU(5) is often included to
obtain correct mass relations for the first and second families of
d-type quarks and leptons. When one adds these higher-dimensional
Higgs representations one must verify the cancellation of
anomalies associated to their fermionic partners. One possible
choice, free of anomalies, include both $45,\overline{45}$
representations to cancel anomalies. We review the necessary
conditions for the cancellation of anomalies and discuss the
different possibilities for supersymmetric SU(5) models.
Alternative anomaly-free combinations of Higgs representations,
beyond the usual vectorlike choice, are identified, and it is
shown that their corresponding $\beta$ functions are not
equivalent. Although the unification of gauge couplings is not
affected, the introduction of multidimensional representations
leads to different scenarios for the perturbative validity of the
theory up to the Planck scale. We study the effect on the
evolution of the gauge coupling up to the Planck scale due to the
different sets of fields and representations that can render an
anomaly-free model.
\end{abstract}

\maketitle




\section{INTRODUCTION}

\textbf{Anomalies in gauge theories}.
    The need to require anomaly cancellation in any gauge theory stems from the fact that their presence destroys the quantum consistency of the theory.
     It turns out that all one needs to calculate or identify the anomaly is the triangle diagrams.
    For a given representation of a gauge group $G$, the anomaly can be written as
\begin{equation}
  A(D)d^{abc}\equiv Tr\left[ \left\lbrace  T_a^{D_i},T_b^{D_i}\right\rbrace T_c^{D_i} \right],
\end{equation}
where $T_a^{D_i}$ are the generators of the gauge group $G$ in the
representation $D_i$, and $d^{abc}$ denotes the anomaly associated
to the fundamental representation \cite{Terning:2006bq}.

    For a representation $R$ that is the direct sum or tensor product of two representations  the anomaly coefficient satisfy
\begin{equation}
\begin{array}{rcl}
  A_{R}=A(R)&=&A(R_{1}\oplus R_{2})=A(R_{1})+A(R_{2}) ,\label{directsum}   \\
  A_{R}=A(R)&=&A(R_{1} \otimes R_{2})=D(R_{1})A(R_{2})+
    D(R_{2})A(R_1), \label{tensorproduct}\end{array}
   \end{equation}
    respectively, with $D(R_{i})$ denoting the dimensions of representation
    $R_{i}$.

 \textbf{Anomaly coefficients}.
    The anomaly coefficients $A(D)$ for most common representations are 
 known in the literature
     \cite{Terning:2006bq}. We have extended these results to include higher-dimensional
representation, with some of them shown in Table
\ref{higherCoeff}.

\begin{table}[ht]\label{higherCoeff}
 \centering
\includegraphics[width=12cm]{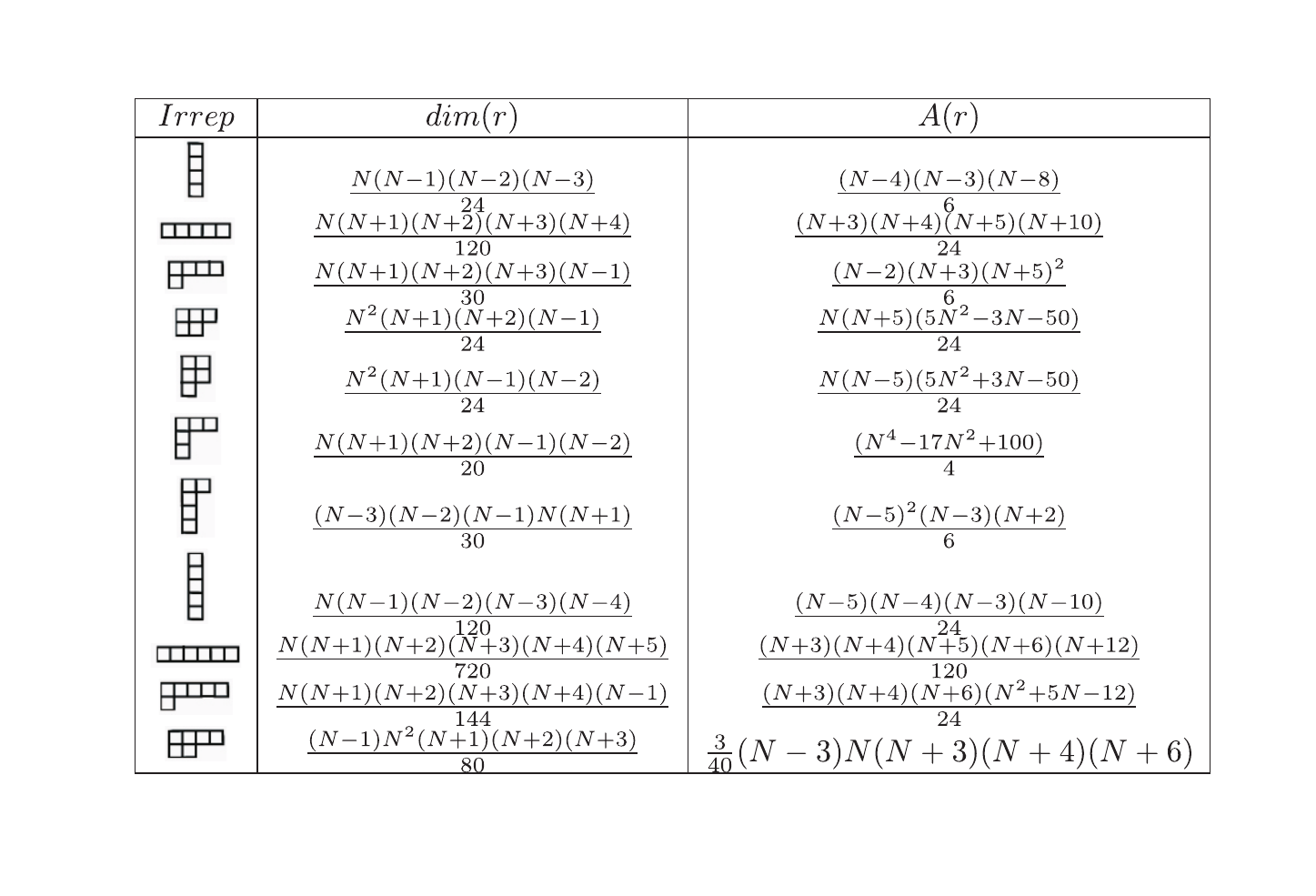}
\caption{Dimension and anomaly coefficients for higher-dimensional
representation of SU(N).}
\end{table}

\begin{table}[ht]
  \caption{Dimension and anomaly coefficients for different representations of SU(5).}
  \label{tabla:SU(5)}
  $\begin{array}{|c|c|c|c|c|}\hline
  Repr.   &Multiplete &dim(r) & A(r)&2T(r)\\
  \hline
  \left[5\right]&(0,0,0,0)&1&0&0\\
  \left[1\right]&(1,0,0,0)&5&1&1\\
  \left[2\right]&(0,1,0,0)&10&1&3\\
  \left[1,1\right]&(2,0,0,0)&15&9&7\\
  Ad &&24&0&10\\
  \left[4,1\right]&(1,0,0,1)&24&0&10\\
  \left[1,1,1\right]&(3,0,0,0)&35&44&28\\
  \left[2,1\right]&(1,1,0,0)&40&16&22\\
  \left[3,1\right]&(1,0,1,0)&45&6&24\\
  \left[2,2\right]&(0,2,0,0)&50&15&35\\
  \hline
  \end{array}$
\end{table}

\section{Anomaly cancellation.}
    There are several ways to construct an anomaly free theory:
\begin{itemize}
    \item The gauge group itself is always free of anomalies. This happens,
for instance, for SO(10) but not for SU(5).
    \item The gauge group is a
subgroup of an anomaly free group, and the representations form a
complete representation of the anomaly free group. For instance,
this happens in the SU(5) case for the
$\mathbf{5}+\mathbf{\overline{10}}$ representations, which
together are anomaly free.
    \item The representations appear in
conjugate pairs, i.e., they are vectorlike. This is the most
common choice when the Higgs sector of SGUT is extended.
\end{itemize}

\textbf{Anomaly cancellation with SU(5) representations.}
       Let us consider an SU(5)  SGUT model.
There are three copies of
$\mathbf{\bar{5}}+\mathbf{\overline{10}}$ representations to
accommodate the three families of quarks and leptons. Breaking of
the GUT group to the SM is achieved by including a (chiral) Higgs
supermultiplet in the adjoint representation (\textbf{24}). The
minimal Higgs sector needed to break the SM gauge group can be
accommodated with a pair of $\mathbf{5}$ and $\mathbf{\bar{5}}$
representations. Within this minimal model one obtains the mass
relations $m_{d_i}=m_{e_i}$, which are predicted by the Higgs
sector. One way to solve this problem is to add a \textbf{45}
representation, which couples to the d-type quarks, but not to the
u-type, then one obtains the Georgi-Jarlskog  factor
\cite{Georgi:1979df} needed for the correct mass relation. Most
models that obtain these relations within an extended Higgs
sector, include the \textbf{45} conjugate representation to cancel
anomalies \cite{FileviezPerez:2007nh}. This is however not de only
possibility.

Consider the representations of SU(5) in Table  \ref{tabla:SU(5)}.
Then, for the \textbf{45}, we can write down the following
anomaly-free combinations:
\begin{eqnarray}
  A(\mathbf{45})+A(\bar{\mathbf{45}})=0,&\nonumber\\
A(\mathbf{45})+nA(\bar{\mathbf{5}})+n'A(\bar{\mathbf{10}})=0,& \nonumber\\ 
 A(\mathbf{45})+A(\bar{\mathbf{15}})+3A(\mathbf{5})=0,&\nonumber
\end{eqnarray}
with  $n$  and  $n'$  integers $\geq 0$ and $n+n'=6$. These are
clearly non-equivalent models, with different physical
consequences.

\section{Gauge coupling unification.}
     The 1-loop $\beta$ functions for a general SUSY theory with gauge group $G$ and matter field appearing in chiral supermultiplets are given by
$\beta=\sum_R T_R - 3C_A$,
 where $T_R$ denotes the index for the
representation $R$, and $C_A$ the quadratic Casimir invariant for
the adjoint representation.
      From MSSM-functions, the 1-loop RGEs for the gauge couplings are
\begin{equation}\label{oneloopRGE}
   \frac{d\alpha_i}{dt}=\beta_i \alpha_i^2,\qquad \beta_i=\left(%
\begin{array}{c}
  33/5 \\
  1 \\
  -3 \\
\end{array}%
\right)+\beta^X.
\end{equation}
where $ \beta^X=\sum_{\Phi} T(\Phi)$ are the contributions of the
extensions of the MSSM \cite{Hempfling:1995rb} and the sum is over
all SU(5) additional multiplets $\Phi$.

We are interested in evaluating the effect of the different
representations in the running from $M_{GUT}\sim 2\times 10^{16}
GeV$ up to the Planck scale, and also in finding which
representations are perturbatively valid up to the Planck scale.
The unified gauge coupling obeys the 1-loop RGE
\begin{equation}
      \mu \frac{d
\alpha_5^{-1}}{d\mu}=\frac{-\beta}{2\pi}=\frac{3-\beta^X}{2\pi}.
\end{equation}

The one loop $\beta$ functions for some interesting anomaly-free
combinations are found to be:

\begin{equation}\label{betafuntions}
    \begin{array}{rcc}
      \beta^X(\mathbf{45}+\bar{\mathbf{45}} )& = & 24, \\
      \beta^X(\mathbf{45}+ 6 (\bar{\mathbf{5}})) & = & 15, \\
      \beta^X(\mathbf{45}+ 6 (\bar{\mathbf{10}})) & = & 21, \\
       \beta^X(\mathbf{45}+ \bar{\mathbf{15}}+2 (\mathbf{10})+\mathbf{5}) & = & 19,\\
            \beta^X(\mathbf{50}+\bar{\mathbf{40}}+\mathbf{5}) & = & 29 .\\
    \end{array}\nonumber
\end{equation}

\begin{figure}[ht]
\centering
 \caption{Evolution of the gauge coupling for free
anomaly combinations listed above, up to the Planck scale . }
\includegraphics[height=0.4\textheight]{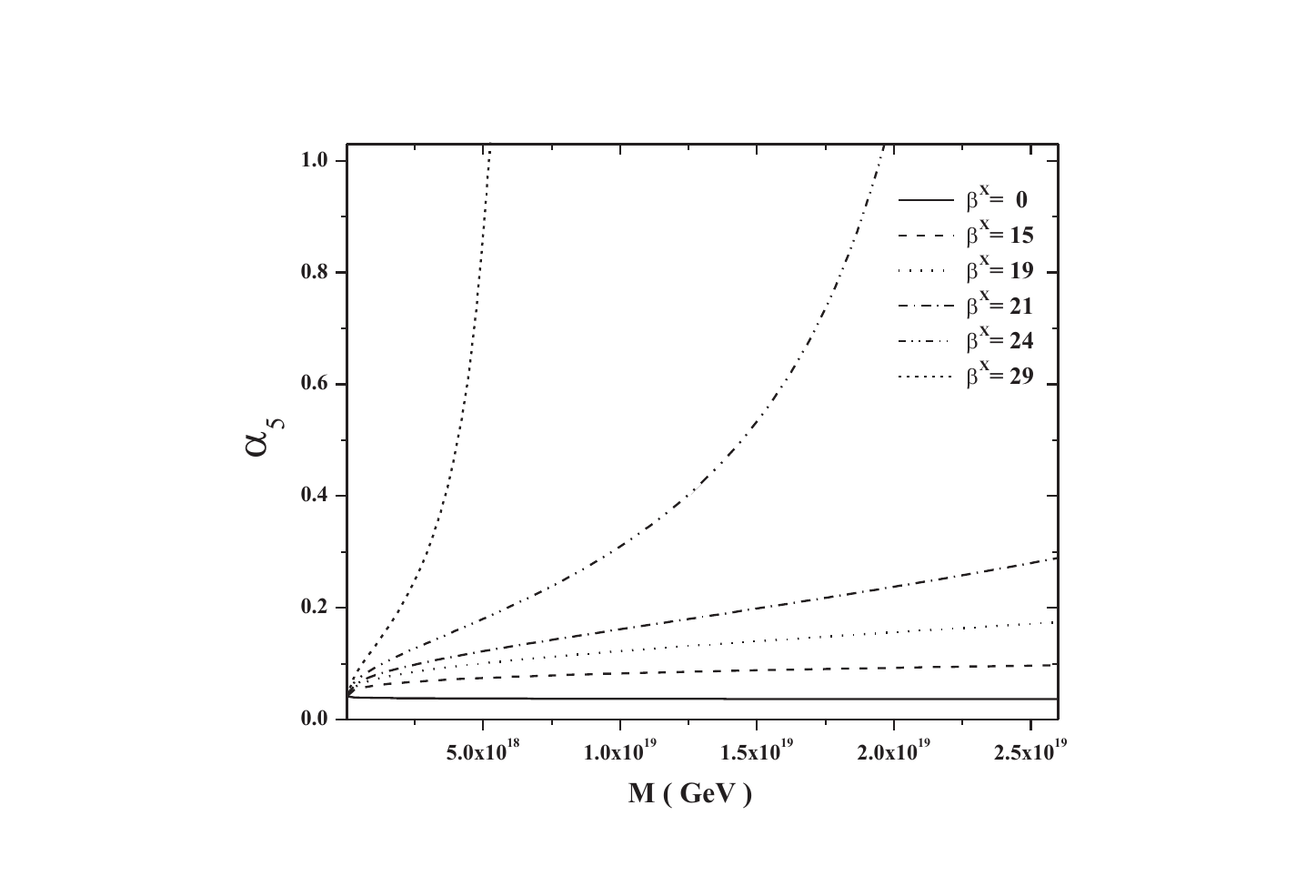}\label{graficaBetas}
\end{figure}

      As it is shown in the figure, the model with $\beta^X= 29$ induces a running of the gauge coupling that blows at the scale
 $M=6.61\times 10^{18}$ GeV, while for $\beta^X = 24$ this happens at $M=2.63\times 10^{19}$ GeV. Models with $\beta^X=15,19,21$
  are found to evolve safely even up to the Planck scale.

\section{Yukawa and gauge couplings.}

      It is also interesting to consider the RGE effect associated with the Yukawa couplings that involve the Higgs representations.
 In order to do this we shall consider the 2-loop $\beta$ functions for the gauge coupling, but will keep only the 1-loop RGE for the Yukawa couplings.
Thus, we shall consider the following superpotential for the SU(5)
SGUT model \cite{Hisano:1992jj}. This superpotential involves the
Higgs representations $H(5)$, $H( 5)$ and $\Sigma(24)$,  and the
matter multiplets $\psi(10)$, $\phi(5)$.

\begin{equation}\begin{array}{ccc}
   W &=&  \frac{f}{3}Tr\Sigma^3+\frac{1}{2}f VTr\Sigma^2
+\lambda\bar{H}_{\alpha}(\Sigma^{\alpha}_{\beta}+3V\delta^{\alpha}_{\beta})H^\beta \\
 && +\frac{h^{ij}}{4} \varepsilon_{\alpha\beta\gamma\delta\epsilon}\psi^{\alpha\beta}_{i}\psi^{\gamma\delta}_{j}H^{\epsilon}
 +\sqrt{2}f^{ij}\psi^{\alpha\beta}_{i}\phi_{j\alpha}\bar{H}_{\beta}, \\
\end{array}
\end{equation}

Using the RGEs expressions for a general supersymmetric model
\cite{Martin:1993zk} we obtain the 1-loop RGE for the Yukawa
parameters \cite{Hisano:1992jj}
\begin{equation}\label{RGE SUSY SU(5)}
\begin{array}{ccc}
  \mu \frac{d\lambda}{d\mu} & = &
  \frac{1}{(4\pi)^2}\left(-\frac{98}{5}g_5^2+\frac{53}{10}\lambda^2
  +\frac{21}{40}f^2+3(h^t)^2\right)\lambda,
\\
 \mu \frac{d f}{d\mu} & = &
  \frac{1}{(4\pi)^2}\left(-30g_5^2+\frac{3}{2}\lambda^2
  +\frac{63}{40}f^2)^2\right)f, \\
\mu \frac{d h^t}{d\mu} & = &
  \frac{1}{(4\pi)^2}\left(-\frac{96}{5}g_5^2+\frac{12}{5}\lambda^2
  +6(h^t)^2\right)h^t, \\
  \end{array}
\end{equation}
while the 2-loop RGE for gauge coupling that we obtain is:
\begin{equation}\label{g5}
    \mu\frac{d
    g_5}{d\mu}=\frac{1}{(4\pi)^2}(-3g_5^{3})+\frac{1}{(4\pi)^4}\frac{794}{5}g_5^{5}-
                \frac{1}{(4\pi)^4}\left\{\frac{49}{5}\lambda^2
                 +\frac{21}{4}f^2+12(h^t)^2\right\}g^3.
\end{equation}

To solve the RGEs we used values of the coefficients $\lambda$,
$h^t$ and $f$ that are themselves safe at the Planck scale. The
parameters used are $M_{GUT}=1.28\times 10^{16}$GeV,
$\alpha(M_{GUT})=0.040$, $h^t(M_{GUT}) = 0.6572$, $\lambda(MGUT) =
0.6024$, and $f (MGUT) = 1.7210$.
\begin{figure}[ht]\label{l7fs}
 \centering
\includegraphics[width=13 cm]{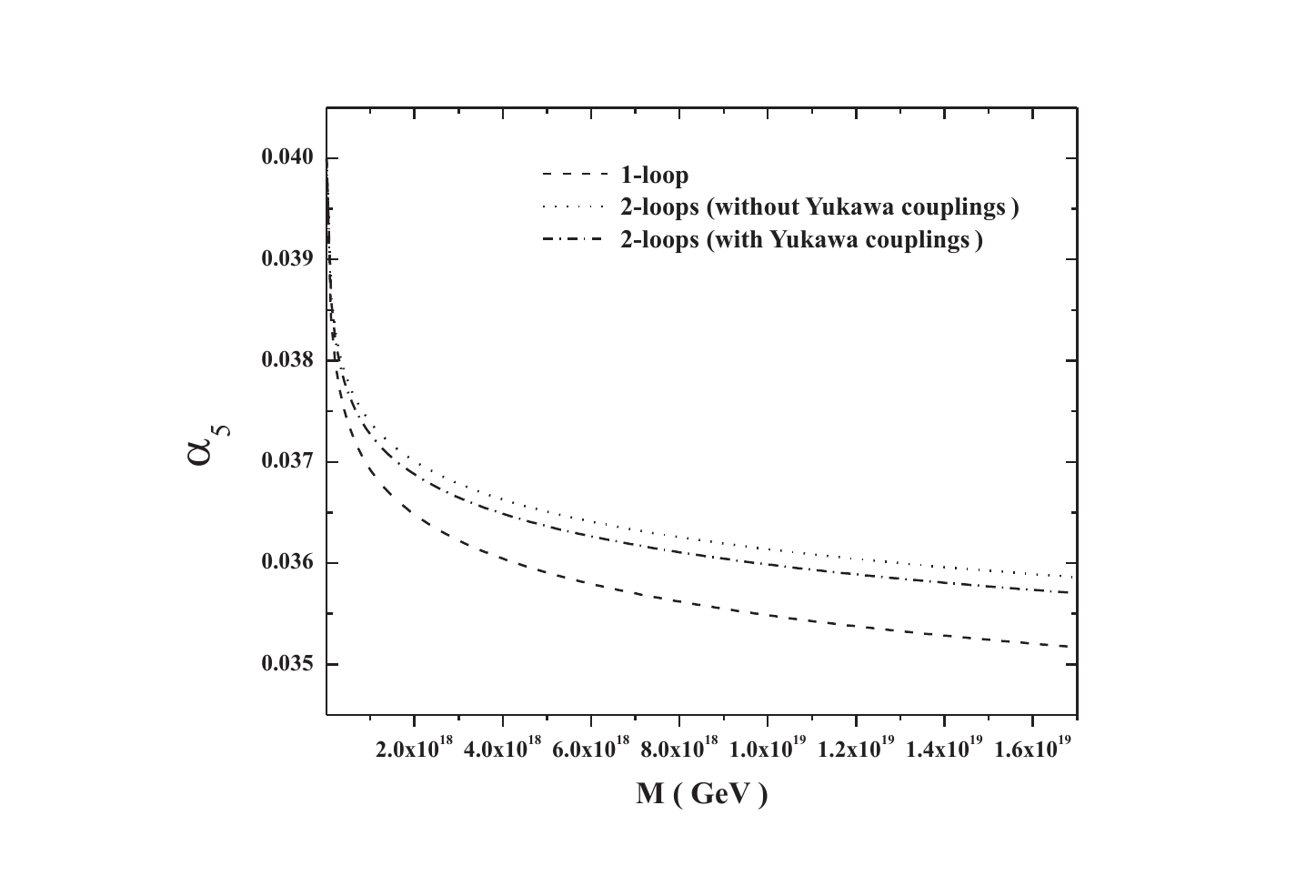}
\caption{Evolution of the unified gauge coupling for three
different cases: i) the 1-loop result, ii) the 2-loop result
without including Yukawas, and iii) the 2-loop result including
the (1-loop) running of the Yukawas.}
\end{figure}

\section{Conclusions}

 We have studied the problem of anomalies in SUSY gauge
theories, in order to search for alternatives to the usual
vectorlike representations used in extended Higgs sector.

The known results have been extended to include higher-dimensional
Higgs representations, which in turn have been applied to discuss
anomaly cancellation within the context of realistic GUT models of
SU(5) type.

 We have succeed in identifying
ways to replace the $\mathbf{\overline{45}}+\mathbf{45}$   models
within SU(5) SGUTs. Then, we have studied the $\beta$ functions
for all the alternatives, and we find that they are not equivalent
in terms of their values. These results have important
implications for the perturbative validity of the GUT models at
scales higher than the unification scale.

 We
have also considered the RGE effect associated with the Yukawa
coupling that involve the additional Higgs representations. We
found that there are appreciable differences for the evolution of
the gauge coupling when going from the 1 to the 2-loops RGE, but
this difference is reduced when one includes the 1-loop Yukawa
couplings at the 2-loop level \cite{Aranda:2009zz,Aranda:2009wh}.


\end{document}